\documentclass[epj]{webofc}
\usepackage[utf8]{inputenc}
\usepackage[varg]{txfonts}   
\usepackage{booktabs}
\usepackage{xcolor}
\definecolor{darkred}{rgb}{0.4,0.0,0.0}
\definecolor{darkgreen}{rgb}{0.0,0.4,0.0}
\definecolor{darkblue}{rgb}{0.0,0.0,0.4}
\usepackage[bookmarks,linktocpage,colorlinks,
    linkcolor = darkred,
    urlcolor  = darkblue,
    citecolor = darkgreen]{hyperref}
\usepackage{subfigure}
\wocname{EPJ Web of Conferences}
\woctitle{Lattice2017}
\begin{document}
%
\selectlanguage{english}
\title{Testing algorithms for critical slowing down}
\author{%
\firstname{Guido} \lastname{Cossu}\inst{1}\fnsep\thanks{speaker, \email{guido.cossu@ed.ac.uk}} \and
\firstname{Peter} \lastname{Boyle}\inst{1} \and
\firstname{Norman} \lastname{Christ}\inst{2} \and
\firstname{Chulwoo} \lastname{Jung}\inst{3} \and
\firstname{Andreas} \lastname{J\"uttner}\inst{4} \and
\firstname{Francesco} \lastname{Sanfilippo}\inst{5} 
}
\institute{%
School of Physics and Astronomy, The University of Edinburgh, Edinburgh EH9 3JZ, United Kingdom
\and
Physics Department, Columbia University, New York, New York 10027, USA
\and
Physics Department, Brookhaven National Laboratory, Upton, New York 11973, USA
\and
School of Physics and Astronomy, University of Southampton, Southampton SO17 1BJ, United Kingdom
\and
Istituto Nazionale di Fisica Nucleare, Sezione di Roma Tre, Via della Vasca Navale 84, I-00146 Rome, Italy
}
\abstract{%
  We present the preliminary tests on two modifications of the Hybrid Monte Carlo (HMC) algorithm. Both algorithms are designed to travel much farther in the Hamiltonian phase space for each trajectory and reduce the autocorrelations among physical observables thus tackling the critical slowing down towards the continuum limit. We present a comparison of costs of the new algorithms with the standard HMC evolution for pure gauge fields, studying the autocorrelation times for various quantities including the topological charge.
}
\maketitle
\section{Introduction}\label{intro}

One of the frontiers of recent Lattice QCD studies is providing precise, sub-percent, Standard Model predictions in the heavy quark sector to compare against the experiments. 
Accurately including dynamical heavy quarks with mass approaching several GeV in the discretised theory simulations requires fine lattice spacings, below 0.05 fm, approaching the continuum limit. 

The continuum limit of a discretised theory is described by a critical point of second order where all the correlation lengths and the correlation times diverge with some power of the inverse lattice spacing, $\tau_{\rm int} \propto a^{-\varepsilon}$. This problem is typically referred as critical slowing down. In general the (integrated) autocorrelation time is going to depend on the algorithm used to generate the Markov chain and on the observable under study. 
For the Langevin algorithm it has been proven \cite{ZinnJustin:1986eq} that this power is universal, $\varepsilon =2$. For the Hybrid Monte Carlo scheme (HMC, sometimes referred as Hamiltonian Monte Carlo)~\cite{Duane:1987de} we do not know the actual exponents and there are claims that these may not be universal since the algorithm is non-renormalisable for the $\phi^4$ theory \cite{Luscher:2011qa}.  

Critical slowing down depends on how much the observable will couple to the low modes of the evolution algorithm in the configuration space. Some observables like the topological charge will be heavily affected by the increasing autocorrelation time. 
The cost to generate ensembles whose sampling is in practice ergodic will increase accordingly. 

 There are several studies in the literature on how to reduce the cost of configuration generation for theories close to the continuum limit \cite{Luscher:2011kk,PhysRevD.90.074502,Schaefer:2010hu,Meyer:2006ty}. In these proceedings we discuss two modifications of the HMC algorithm that have the potential to reduce the total cost of the expensive dynamical simulations with fermions. The first modification, called Riemannian Manifold HMC (RMHMC), modifies the kinetic term of the HMC Hamiltonian to speed up the evolution of the slowest modes. The second, referred as Look Ahead HMC (LAHMC), drops the detailed balance requirement in favour of a generalized principle with the effect of an increased trajectory length.

The first step is to show effective reductions of the autocorrelation times and related algorithm cost in the simple cases of SU(3) pure gauge theory and $CP^N$ models, that already exhibit critical slowing down \cite{Schaefer:2010hu,Campostrini:1992ar,DAdda:1978vbw,Witten:1978bc,DelDebbio:2004xh}.  

\section{Description of the algorithms}\label{descr}

In this section we describe the details of the algorithms starting with a brief summary of the HMC to fix the notation.

\subsection{Markov Chain Monte Carlo}\label{S:MCMC}

The basic problem of simulating a quantum field theory is that we have to sample field configurations distributed according to some target probability $p(x) \propto \exp (-S(x))$ where $S(x)$ is the corresponding action functional of the field $x$. If we are able to generate a Markov chain sequence with transition probabilities $T(x^\prime|x)$ from the state $x$ to state $x^\prime$ that satisfies
\begin{equation}
\int p(x) T(x^\prime|x){\rm d}x = p(x^\prime)
\end{equation}
i.e. $p(x)$ is a \emph{fixed point} of the transition matrix\footnote{In a discrete state space the target distribution is an eigenstate of the transition matrix.}, then the expectation value of an observable can be computed by simply averaging the measurements over the generated ensemble. For this purpose, a sufficient but not necessary condition is for $T(x^\prime|x)$ to satisfy the detailed balance relation:
\begin{equation}\label{eq:det-balance}
p(x) T(x^\prime|x) = p(x^\prime) T(x|x^\prime).
\end{equation}
i.e. that for the pair $x$, $x^\prime$, the probability of going from state $x$ to $x^\prime$ is the same as the probability
of reaching state $x$ from $x^\prime$. Equation (\ref{eq:det-balance}) automatically implies the fixed point condition, but introduces a random walk component in the HMC. 
This relation is satisfied by the accept-reject step in the HMC.

\subsection{Hybrid Monte Carlo}\label{S:HMC}

The Hybrid Monte Carlo scheme \cite{Duane:1987de} for the generation of correctly distributed configurations for a field $\phi$ extends the space with a fictitious time coordinate and generates new configurations according to the distribution $q(\phi)$:
\begin{equation}
q(\phi) \propto \exp(-H(\phi)) \qquad H(P, \phi) = \frac{1}{2} P^TP + S(\phi)
\end{equation}
where $P$ are the conjugate momenta for $\phi$ in the Hamiltonian evolution and it enters as a trivial Gaussian factor. An HMC update step can be easily described as an application of a sequence of operators. In order to fix our notation these operators are 
\begin{itemize}
\item $L(\epsilon, M)$, any integrator that is area preserving and reversible. Runs the integration for $M$ steps of size $\epsilon$.
\item $F$, momenta sign-flip
\item $R(\alpha)$, randomization of the momenta by Gaussian noise, $\alpha \in [0,1]$. The parameter $\alpha$ is the mixing between the new random momenta $P_{\rm new}$ and the old ones $P_{\rm old}$, i.e. $R(\alpha) = \sqrt{\alpha} P_{\rm new} + \sqrt{(1-\alpha)} P_{\rm old}$  
it is typical to set $\alpha = 1$, i.e. a complete randomization of momenta.
\end{itemize}

Using these operators we can write the HMC in a compact form as follows. The $x^{(t,n)}$ stands for the configuration state $(P,\phi)$ at the point $t$ of the Markov chain and at the step $n$ of the evolution sequence to generate the new Markov chain state. We have 3 sub-states $n = 0,1,2$ in the HMC for the generation of a new state.   
\begin{enumerate}
\item $x^\prime = FL x^{(t,0)}$, where $x^{(t,0)}$ is the initial state. Then compute the acceptance probability using the Metropolis step and $x^{(t,1)} = x^\prime$ if accepted, otherwise $x^{(t,1)} = x^{(t,0)}$. 
\item Flip the momenta. $x^{(t,2)} = F x^{(t,1)}$.
\item Randomize momenta. The new element of the Markov chain is $x^{(t+1,0)} = R(\alpha) x^{(t,2)}$.
\end{enumerate}

The $FL$ step is necessary for the HMC in order to cancel the contribution of forward and backward transition probabilities in the Metropolis-Hastings acceptance probability of state $x^\prime$, $\pi(x | x^\prime)$
\begin{equation}\label{eq:Metropolis}
  \pi(x | x^\prime)  = \min \Bigl(1, \frac{p(x^\prime)}{p(x)}\frac{T(x | x^\prime)}{T(x^\prime | x)} \Bigr)
\end{equation}
since $(FL)^{-1} = L^{-1}F = FL$.

\subsection{Riemannian Manifold HMC}\label{S:RMHMC}

One issue in the evolution induced by HMC algorithm is that in general, in the configuration space we have fast modes of the transition matrix evolving quickly and some slow modes that will take longer time to complete a cycle and decorrelate. Fast and slow corresponds to large and small gauge forces in the HMC and the speed refers to the MD time required to evolve over a full period.
The former will limit the integration step size from above, otherwise the integrator becomes unstable. The divergence of the ratio of the highest and the lowest frequencies determines the degree of critical slowing down. In light of this observation, an obvious solution is trying to make all the modes to evolve with the same frequency. 

We now follow the ideas of Duane et al.~\cite{Duane:1988vr,Duane:1986fy} on Fourier acceleration of slow modes in the evolution of the Markov chain. This leads to a non-separable Hamiltonian in the HMC which we integrate with the generalised leapfrog integrator~\cite{Leimkuhler:835066} as suggested by Girolami and Calderhead~\cite{RSSB:RSSB765}. The resulting algorithm acknowledges that the Hamiltonian manifold where the evolution takes place is not flat and modifies the kinetic term with a metric accordingly, e.g. with the inverse of the Laplacian operator.

Consider the simplest example of acceleration of the slow modes in a free scalar field theory with mass $m$ that have characteristic frequencies $\omega^2(p)= m^2 + p^2$. Let us study the evolution of the modified Hamiltonian with a new parameter $M$~\cite{Duane:1986fy}:
\begin{equation}
H(\pi, \phi) = \frac{1}{2} \pi^T (M^2 -\partial^2)^{-1}\pi + S(\phi) =\frac{1}{2} \pi^T \mathcal{M}^{-1}\pi + S(\phi) , \qquad S(\phi) = \frac{1}{2}\phi(m^2 - \partial^2)\phi.
\end{equation}
 The Hamiltonian $H$ has now characteristic frequencies:
\begin{equation}
\omega^2(p)= \frac{m^2 + p^2}{M^2+p^2} 
\end{equation}
that we can tune in order to have a function independent on $p$ and have perfect acceleration. The inverse  in the new kinetic term is easily written in the momentum space, hence the name Fourier acceleration. The acceleration parameter for interacting theories will have to be determined, but a good choice is the renormalized mass of the scalar field \cite{Duane:1988vr}.

Gauge theories pose several complications in this context, see \cite{Duane:1986fy}. First we need a covariant version of the kinetic term, that we are going to denote as the kinetic kernel. There are many possible choices for the kinetic kernel, in this proceedings we are going to use the covariant Laplacian, that for fields in the adjoint representation is:
\begin{equation}
\nabla^2 \phi(x) = \sum^d_\mu \Bigl[ U_\mu(x) \phi(x+\mu) U^\dagger_\mu(x) + U^\dagger_\mu(x-\mu) \phi(x-\mu) U_\mu(x-\mu) - 2 \phi(x) \Bigr]  
\end{equation}
for its discretised version, $U(x)$ being the gauge field. In the RMHMC Hamiltonian, the operator $\mathcal{M}$ acting on fields in the adjoint representation is written as
\begin{equation}
\mathcal{M} \pi(x) = (1 - \kappa) \pi(x) - \frac{\kappa}{4 d} \nabla^2 \pi(x) 
\end{equation}
where $d$ is the dimensionality of the theory and we get maximal decoupling (in the free field theory) for $\kappa \rightarrow 1$. Zero modes of the Laplacian operator constrain $\kappa \neq 1$.

The resulting Hamiltonian equations are now non-separable and as such the standard St\"ormer-Verlet leapfrog integration algorithms (and their higher order versions) are not useful any more because they are not reversible and the Jacobian of the transformation does not have unit determinant, i.e. it is not volume preserving. 

Leimkuhler and Reich~\cite{Leimkuhler:835066} proposed a generalised leapfrog scheme that is reversible. In terms of the gauge field theory variables $(\pi, U)$ it is written
\begin{align}\label{eq:implicit}
  \pi^{n+\frac{1}{2}} &= \pi^{n} - \frac{\epsilon}{2}\frac{\delta H}{\delta U}(U^n, \pi^{n+\frac{1}{2}})\\
U^{n+1} &= \exp \Bigl( \frac{\epsilon}{2}\Bigl[\frac{\delta H}{\delta \pi}(U^n, \pi^{n+\frac{1}{2}}) +\frac{\delta H}{\delta \pi}(U^{n+1}, \pi^{n+\frac{1}{2}})    \Bigr] \Bigr) U^n \\
\pi^{n+1} &= \pi^{n+\frac{1}{2}} - \frac{\epsilon}{2}\frac{\delta H}{\delta U}(U^{n+1}, \pi^{n+\frac{1}{2}})
\end{align}
We can eventually concatenate more than one of these steps by symmetrising this sequence to make it reversible (new implicit steps in the update of $\pi$ will appear). The implicit equations can be solved iteratively. 

The equation (\ref{eq:implicit}) is general, valid even when fermions are included. In particular notice that the implicit evaluation of the force term would not require any additional computation of the $\frac{\delta H}{\delta U}(U^{n})$ force term during the iterative process.

The modification of the kinetic term introduces the unwanted determinant of the matrix $\mathcal{M}$ in the path integral. To cancel this term we introduced a new set of auxiliary bosonic fields $\phi_{\mu}$ in the adjoint representation with action ${\rm Tr}[\phi_{\mu}\mathcal{M}\phi_{\mu} + \phi^2]$, see \cite{Duane:1986fy,Duane:1988vr}.

We implemented the RMHMC with implicit versions of the leapfrog and the minimum norm 2nd order integrators in Grid \cite{Boyle:2016lbp}. Furthermore, the code is ready for simulations with more complex actions.

\section{Look Ahead HMC}

It has already been shown that increasing the trajectory length in the HMC algorithm can improve the autocorrelation times, for the same or reduced cost \cite{Meyer:2006ty,Schaefer:2010hu}. The basic idea is to move further in the integration space before randomizing the momenta and let the slow modes complete their cycle. 

In order to suppress the random walk behaviour induced by the introduction of the accept-reject step needed to ensure the detailed balance it has been suggested to modify the Metropolis accept-reject step \cite{Sohl-Dickstein:2014,Campos2015JCoPh}.
The resulting probability distribution satisfies the fixed point equation but drops the detailed balance condition. The final effect would be a reduction of the rejection rate for the same step size and effective increase of the trajectory length, and eventually an optimization of the costs.

Using the notation introduced in section~\ref{S:HMC} the new LAHMC algorithm \cite{Sohl-Dickstein:2014,Campos2015JCoPh} consists in the following steps:
\begin{itemize}
\item Repeat the integration accepting with the modified probabilities $\pi_{L^K}(x^{(t,0)} )$ for a maximum number $K$ of times.
  \begin{equation}
    x^{(t,1)} = 
    \left\{
      \begin{tabular}{l l l }
        $L x^{(t,0)}$    && prob $\pi_{L^1}(x^{(t,0)} )$   \\
        $L^2 x^{(t,0)}$  && prob $\pi_{L^2}(x^{(t,0)} )$    \\
        $\cdots$ \\
        $L^K x^{(t,0)}$  && prob $\pi_{L^K}(x^{(t,0)} )$    \\
        $F x^{(t,0)}$    && prob $\pi_F(x^{(t,0)} )$ 
      \end{tabular}
    \right.
  \end{equation}
\item Randomize momenta $x^{(t+1,0)} = R(\alpha) x^{(t,1)}$.
\end{itemize}

The transition probabilities $x \rightarrow L^a(x)$ are defined in order to satisfy the fixed point equation and are:
\begin{equation}\label{eq:LAHMCprob}
  \pi_{L^a}(x) = \min \Bigl[1- \sum_{b<a} \pi_{L^b}(x), \frac{p(FL^ax)}{p(x)}\Bigl(1-\sum_{b<a} \pi_{L^b}(FL^ax)\Bigr) \Bigr]
\end{equation}
where $p(x)$ is the same probability function as in equation (\ref{eq:Metropolis}). These transition probabilities do not satisfy the detailed balance relation but a more generalized set of identities, called generalized detailed balance \cite{Campos2015JCoPh}. The target distribution is still a fixed point of the evolution algorithm. Please refer to equations (28)-(33) of \cite{Sohl-Dickstein:2014} for a straightforward demonstration of this fundamental result. Notice that for $K=1$ the LAHMC reduces to the usual HMC. The practical implementation of equation (\ref{eq:LAHMCprob}) involves a solution of an implicit set of equations. 

The LAHMC has been implemented in Grid \cite{Boyle:2016lbp} and the two new parameters are $K$, the maximum number of repetitions, and $\alpha$, the mixing factor in the momenta generation. Prolonging the integration has a high chance of decreasing the rejection rate since we are working with symplectic integrators that fluctuate around the correct integration surface during the evolution.

\section{Preliminary numerical results}\label{S:Results}

We now present some of our preliminary results on the efficiency and cost of these two algorithms in the case of the pure gauge SU(3) and the $CP^N$ model with $N=10$. 

The gauge action is the standard Wilson action with $\beta=6.2$ ($a = 2.9$ Gev$^{-1}$ = 0.068 fm). The lattice size is $L^4= 16^4$. The integrator is always the 2nd order minimum norm \cite{Takaishi:2005tz}, in its standard or implicit version (RMHMC). We define $N_{MD}$ as the number of molecular dynamics steps in a trajectory and $\tau$ as the trajectory length. We measure the observables every 5 generated configurations and all the autocorrelation times are consistently reported in these units. For the LAHMC we tested different values of $\alpha$ but these had no appreciable effect and are not included in the plots due to larger errors. The parameters of the run LA1 were chosen to reduce the cost and Ref0 (standard HMC) to match the acceptance rate. 

Runs for the $CP^N$ model are performed at two different $\beta$ and volumes: $\beta=0.7, L=42$ and $\beta=0.9, L=90$ respectively. Measurements are performed after every trajectory.

\begin{table}[t]\label{tab:data}
  \centering
  \small
  \begin{tabular}{ l | l c c c c c c c c l }
    \toprule
    ID & $L$  & $\beta$ &  $N_{MD}$  & $\tau$ & K& $\alpha$ & $\kappa$ & $\langle dH \rangle$ & Acceptance \% & Acc1\\
    \midrule
    Ref0 & 16 & 6.2 & 17 & 1.0 & - & -  & -   &0.087 $\pm$ 0.0028 & 0.83 & -    \\
    LA0 & 16 & 6.2 & 10 & 1.0 & 5 & 0.6 & -   &0.704 $\pm$ 0.0080 & 0.81 & 0.55 \\
    LA1 & 16 & 6.2 & 10 & 1.0 & 5 & 1.0 & -   &0.697 $\pm$ 0.0061 & 0.81 & 0.55 \\
    RM0 & 16 & 6.2 & 30 & 1.0 & - & -   & 0.3 &0.013 $\pm$ 0.0006 & 0.93 & -    \\
    RM1 & 16 & 6.2 & 30 & 1.0 & - & -   & 0.6 &0.017 $\pm$ 0.0006 & 0.93 & -    \\
    RM2 & 16 & 6.2 & 30 & 1.0 & - & -   & 0.8 &0.026 $\pm$ 0.0009 & 0.91 & -    \\
    RM3 & 16 & 6.2 & 30 & 1.0 & - & -   & 0.9 &0.036 $\pm$ 0.0011 & 0.89 & -    \\
    \bottomrule 
  \end{tabular}
  \caption{Parameters for the runs presented in these proceedings. The Acc1 column shows the acceptance rate of only the first trajectory in the LAHMC evolution. Other quantities are defined in the text. $\langle dH \rangle $ is the average energy deviation per trajectory. The run LA0 is not included in the figures.}
\end{table}

The average cost for an observable $\mathcal{O}$ in molecular dynamics steps units per configuration, $C_{\mathcal{O}}$, is defined as:
\begin{equation}
  \label{eq:cost}
  C_{\mathcal{O}} = \frac{N_{\rm traj} N_{MD}}{N_{\rm conf}}\tau_{\rm int}(\mathcal{O})
\end{equation}
where $N_{\rm traj}$ and $N_{\rm conf}$ are respectively the number of trajectories generated (in units of $\tau$) and the number of new configurations generated per ensemble. The integrated autocorrelation time of $\mathcal{O}$ is indicated by $\tau_{\rm int}(\mathcal{O})$. 
For the HMC and the RMHMC case $N_{\rm traj}$ and $N_{\rm conf}$ are the same. This is not the case for the LAHMC algorithm where for each new configuration we can travel more than one trajectory length. $C$ represents the average number of molecular dynamics steps per new configuration. 
Travelling for one molecular dynamics step costs exactly the same for the algorithms -- in terms of computations of the force terms $ \frac{\delta H}{\delta U}(S(U^{n})) $ -- so use this quantity here for a fair comparison (also in view of the application of these algorithms to the more expensive simulations with dynamical fermions). 
In the case of the RMHMC we are not including the cost of the inversion of the Laplacian and the implicit steps. The Laplacian inversion is relatively cheap, converging in O(20) iterations with very mild dependence on $\kappa$. The implicit steps converge in less than 10 iterations and require no additional computations of the force terms relative to the action $S(U)$, very expensive in the case of fermion actions. These steps would be negligible when dynamical fermions at the physical point are included in the simulation, since the fermion force term does not have to be evaluated for each implicit step.

In the SU(3) case the observables are measured after applying some smearing (Wilson Flow, $\tau_{\rm flow} = 2.0$): they are the average action $S(U)$ and $Q^2$, that is proportional to the topological susceptibility. 
Integrated autocorrelation times are estimated following \cite{Wolff:2003sm}. 
The observables for the $CP^N$ model are the zero momentum projected 2-point function, $G_0$, and the topological charge. 

\begin{figure}[ht]
  \centering
  \includegraphics[width=0.45\columnwidth, clip]{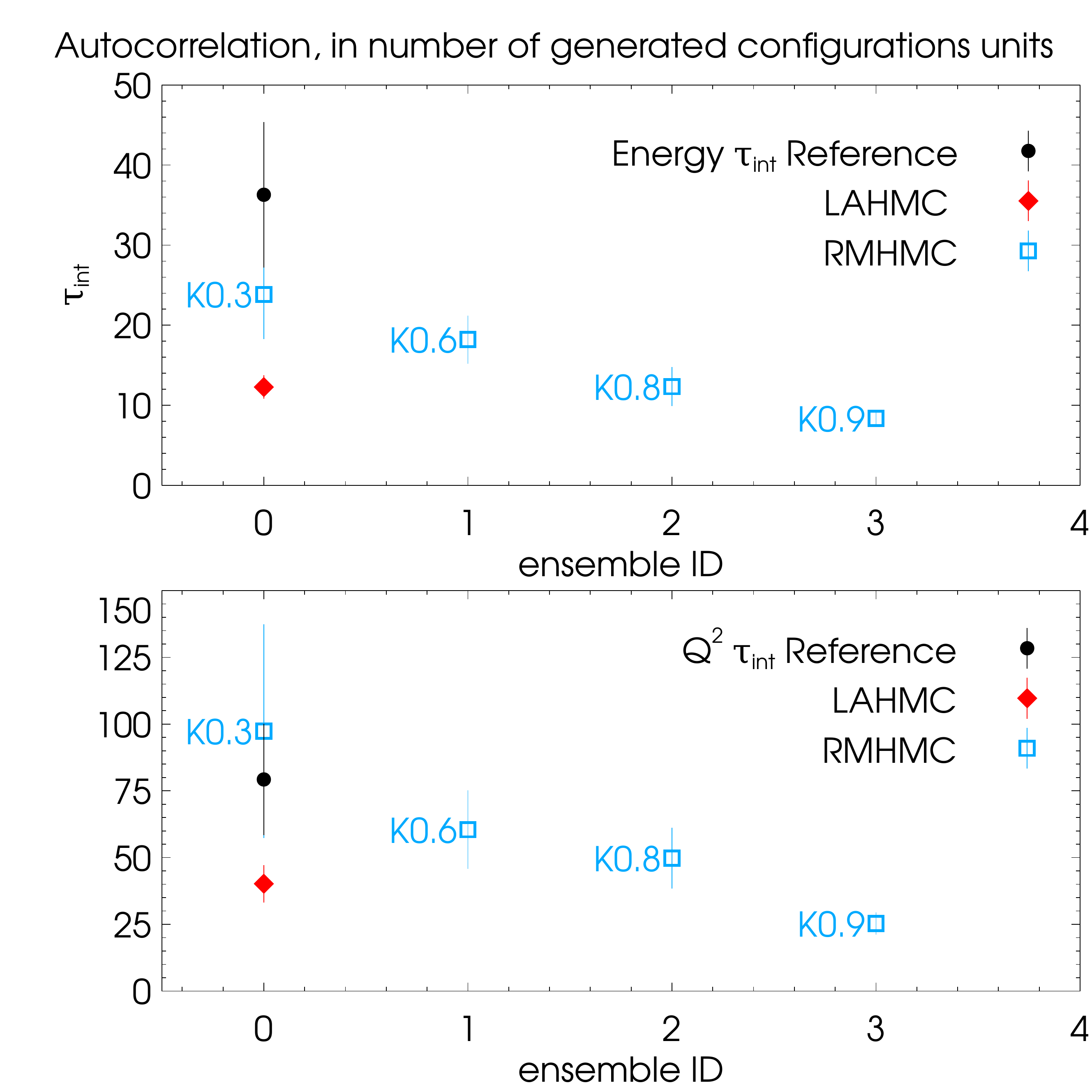}
  \includegraphics[width=0.45\columnwidth, clip]{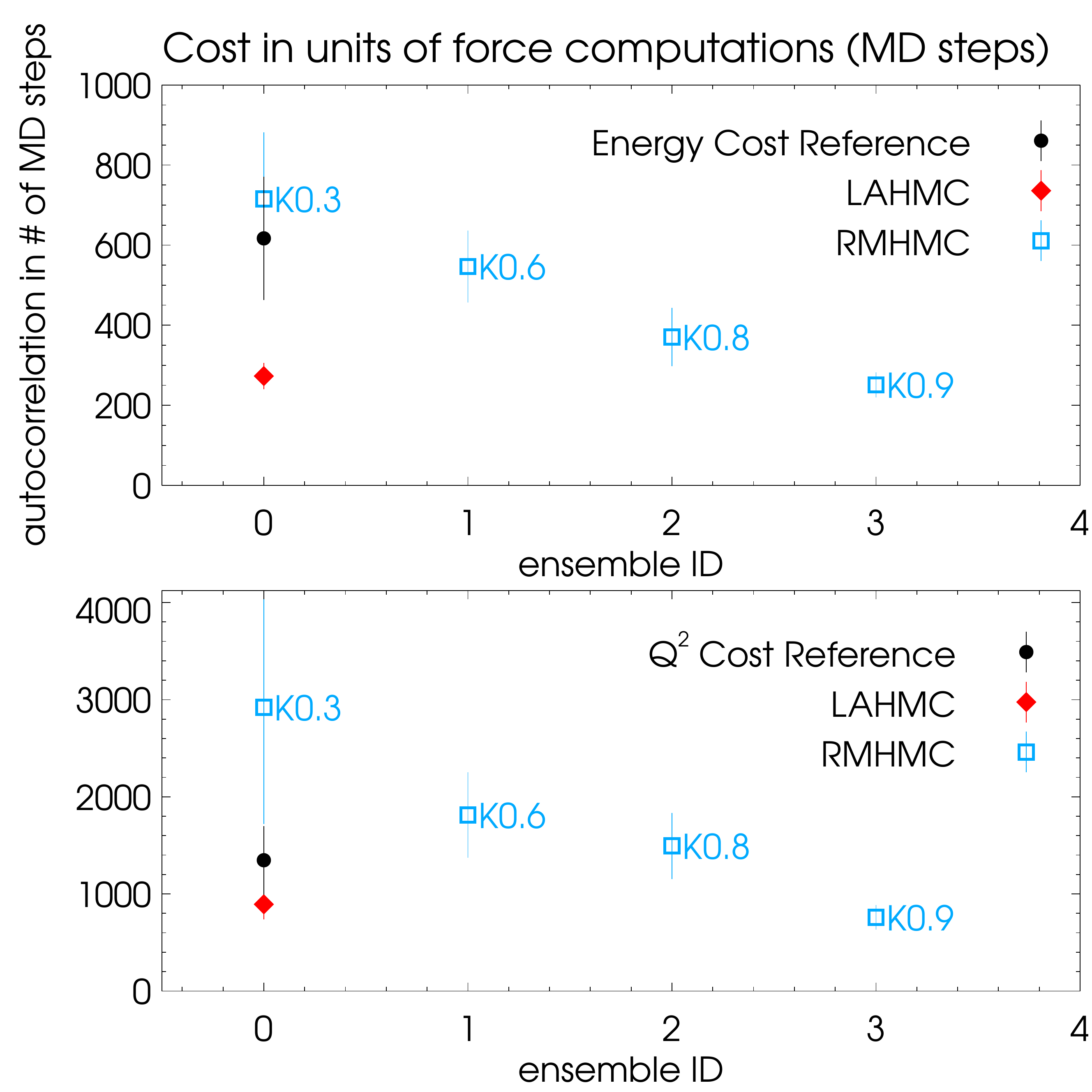}
  \caption{On the $x$-axis the ensemble ID from the first column of table \ref{tab:data}. Left: Integrated autocorrelation times in units of the generated configurations, i.e. measuring every 5 generated configurations. Upper panel show the autocorrelation for average action $F_{\mu\nu} F_{\mu\nu}$ and the lower panel for the topological susceptibility. Right:  Cost estimates in units of the molecular dynamics steps, see equation (\ref{eq:cost}).}
   
  \label{fig-1}
\end{figure}

The results for the absolute autocorrelation time and the cost estimate in the SU(3) pure gauge case are reported in figure \ref{fig-1}. 
The cost estimate shows promising results for both the RMHMC and the LAHMC that are almost equivalent at the maximal acceleration.

\begin{figure}[ht]
  \centering
  \includegraphics[width=0.50\columnwidth, clip]{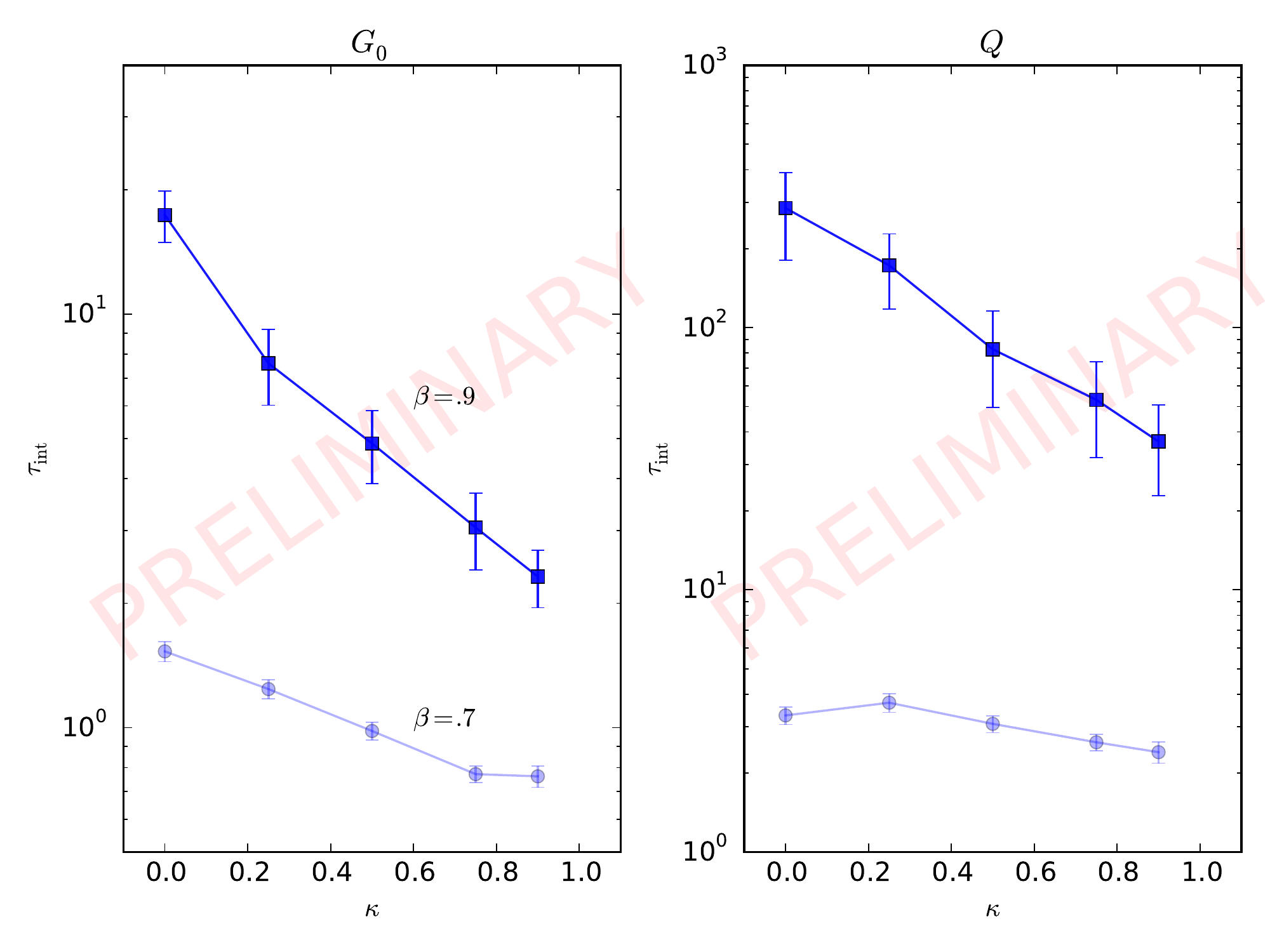}
  \caption{Integrated autocorrelation times for the $CP^N$ model (N=10). Left: autocorrelation for the integrated correlator for two different $\beta$ and its dependence on the acceleration parameter $\kappa$. Right: same analysis for the topological charge.}
  \label{fig-cpn}
\end{figure}

The results for the $CP^N$ model in figure \ref{fig-cpn} are even more striking showing a reduction of one order of magnitude of the autocorrelation times of $G_0$. It is important to notice that the RMHMC algorithm effectiveness is increasing towards the continuum limit $\beta \rightarrow \infty$.

\section{Summary}

We presented our preliminary investigation on two variants of the Hybrid Monte Carlo \cite{Duane:1987de} in order to reduce the cost of configuration generation in the region of the parameter space useful for high precision flavour physics. The Riemannian Manifold HMC (RMHMC) \cite{Duane:1988vr,RSSB:RSSB765,Duane:1986fy} modifies the kinetic term in the HMC Hamiltonian to acknowledge the presence of faster and slower modes during the evolution. The Look Ahead HMC (LAHMC) drops the detailed balance condition in favour of a generalized detailed balance that would reduce the random walk behaviour that the acceptance-reject step introduces and increase the effective trajectory length at a lower cost.

The preliminary results presented in these proceedings show that both algorithms have the potential to reduce the computations required to generate independent configurations and be ergodic. Moreover the two algorithms act in orthogonal directions and thus can be easily combined for even better decorrelation, a possibility to be tested. Another direction of research, once we confirm the viability of these methods, is the combination of these algorithm with other methods proposed in the literature to improve the ergodicity by keeping the costs under control.

We are currently increasing the statistics of our runs and probing the hyper-parameter space. In the pure gauge case we have runs for lattices $L^4=32^4$ for both algorithms at $\beta=6.4$ in order to check the scaling toward the continuum limit as done already for the $CP^N$ case.

\section{Acknowledgements}

The work of GC is supported by STFC, grant ST/L000458/1. 
NC is supported in part by US DOE grant \#DE-SC0011941. 
PB is an Alan Turing Institute Faculty Fellow, acknowledges STFC grants ST/M006530/1, ST/L000458/1, ST/K005790/1, ST/K000411/1, ST/H008845/1, and ST/K005804/1, and Royal Society Wolfson Research Merit Award WM160035. 
CJ is supported in part by U.S. DOE Contract \#AC-02-98CH10886 (BNL). 
AJ is supported by the European Research Council under the European Union’s Seventh Framework Programme (FP7/2007- 2013) / ERC Grant agreement 279757.
The simulations ran on the Cambridge CSD3 Xeon-Phi cluster and the Plymouth University cluster (thanks to Antonio Rago). 
\bibliography{lattice2017}
\end{document}